\documentclass[a4paper,useAMS,usenatbib]{mn2e}

\usepackage{mathptmx}

\usepackage[T1]{fontenc}
\usepackage{ae,aecompl}

\usepackage{graphicx}	
\usepackage{amsmath}	
\usepackage{amssymb}	

\title[Imaging of H~{\sc i} absorption structure at $z = 3.1$]{Imaging of diffuse H~{\sc i} absorption structure in the SSA22 proto-cluster region at $z = 3.1$}

\author[K. Mawatari et al.]{
Ken Mawatari,$^{1}$\thanks{E-mail: mawatari@las.osaka-sandai.ac.jp}
Akio K. Inoue,$^{1}$
Toru Yamada,$^{2}$
Tomoki Hayashino,$^{3}$
\newauthor
Takuya Otsuka,$^{4}$
Yuichi Matsuda,$^{5}$
Hideki Umehata,$^{6,7}$
Masami Ouchi$^{8}$
\newauthor
and Shiro Mukae$^{8,9}$
\\
$^{1}$College of General Education, Osaka Sangyo University, 3-1-1, Nakagaito, Daito, Osaka, 574-8530, Japan\\
$^{2}$Institute of Space Astronautical Science, Japan Aerospace Exploration Agency, Sagamihara, Kanagawa 252-5210, Japan\\
$^{3}$Research Center for Neutrino Science, Graduate School of Science, Tohoku University, Aoba, Aramaki, Aoba-ku, Sendai, Miyagi, 980-8578, Japan\\
$^{4}$Astronomical Institute, Tohoku University, Aoba, Aramaki, Aoba-ku, Sendai, Miyagi, 980-8578, Japan\\
$^{5}$National Astronomical Observatory of Japan, Osawa 2-21-1, Mitaka, Tokyo 181-8588, Japan\\
$^{6}$The Open University of Japan, 2-11 Wakaba, Mihama-ku, Chiba-shi, Chiba, 261-8586, Japan\\
$^{7}$Institute of Astronomy, The University of Tokyo, 2-21-1 Osawa, Mitaka, Tokyo 181-0015, Japan\\
$^{8}$Institute for Cosmic Ray Research, The University of Tokyo, 5-1-5 Kashiwanoha, Kashiwa, Chiba 277-8582, Japan\\
$^{9}$Department of Astronomy, Graduate School of Science, The University of Tokyo, 7-3-1 Hongo, Bunkyo, Tokyo, 113-0033, Japan
}

\date{Accepted XXX. Received YYY; in original form ZZZ}

\pubyear{2016}

\begin{document}
\label{firstpage}
\pagerange{\pageref{firstpage}--\pageref{lastpage}}
\maketitle

\begin{abstract}
Using galaxies as background light sources to map intervening Ly$\alpha$ absorption is a novel approach to study the interplay among galaxies, the circum-galactic medium (CGM), and the intergalactic medium (IGM). Introducing a new measure of $z = 3.1$ H~{\sc i} Ly$\alpha$ absorption relative to the cosmic mean, $\Delta NB497$, estimated from photometric data of star-forming galaxies at $3.3 \la z \la 3.5$, we have made two-dimensional $\Delta NB497$ maps in the $z = 3.1$ SSA22 proto-cluster region and two control fields (SXDS and GOODS-N fields) with a spatial resolution of $\sim 5$\,comoving Mpc. The $\Delta NB497$ measurements in the SSA22 field are systematically larger than those in the control fields, and this H~{\sc i} absorption enhancement extends more than $50$\,comoving Mpc. The field-averaged (i.e., $\sim 50$\,comoving Mpc scale) $\Delta NB497$ and the overdensity of Ly$\alpha$ emitters (LAEs) seem to be correlated, while there is no clear dependency of the $\Delta NB497$ on the local LAE overdensity in a few comoving Mpc scale. These results suggest that diffuse H~{\sc i} gas spreads out in/around the SSA22 proto-cluster. We have also found an enhancement of $\Delta NB497$ at a projected distance $< 100$\,physical kpc from the nearest $z = 3.1$ galaxies at least in the SSA22 field, which is probably due to H~{\sc i} gas associated with the CGM of individual galaxies. The H~{\sc i} absorption enhancement in the CGM-scale tends to be weaker around galaxies with stronger Ly$\alpha$ emission, which suggests that the Ly$\alpha$ escape fraction from galaxies depends on hydrogen neutrality in the CGM. 
\end{abstract}




\section{Introduction}

Galaxy formation occurs through assembly of gaseous matter followed by star-formation in the dark matter large-scale structure (``Cosmic Web''). The dark matter structure formation, which is strictly governed by theory of the gravity, has been well predicted in the many theoretical and numerical works (e.g., \citealt{WhiteRees78,Davis+85,Springel+05,Mo+10}). The star-formation and stellar component of galaxies are the most common observable and routinely studied (e.g., \citealt{Santini+09,Daddi+10,Genzel+10,Smit+12,Madau+14}). On the other hand, the assembly of gaseous matter from the ``Cosmic web'' of the intergalactic medium (IGM) onto galaxies is still a rarely explored territory. The kinetic, radiative, and chemical feedbacks from galaxies play an important role in the IGM metal enrichment, as well as in the formation of galaxies themselves (e.g., \citealt{Scannapieco+02,Tremonti+04,Simcoe+06,AguirreSchaye07,Kobayashi+07,Cen+11}). Understanding the interplay of gas between galaxies and the ``Cosmic Web'', especially behavior of gas in the circum-galactic scale (circum-galactic medium; CGM), is essential to develop a complete galaxy formation and evolution theory. 

The relation between galaxies and neutral hydrogen (H~{\sc i}) gas in the IGM/CGM has been investigated spectroscopically with the Ly$\alpha$ absorption imprinted in spectra of background QSOs (e.g., \citealt{Adelberger+03,Rudie+12,Rakic+12,Turner+14,Crighton+15,Mukae+16}) and star-forming galaxies (e.g., \citealt{Adelberger+05,Steidel+10,Lee+16,Mawatari+16}). Some authors recently discovered an enhancement of the H~{\sc i} absorption associated with proto-clusters at $z = 2$ -- $3$ by using QSO sight-lines \citep{Cai+16} or by stacking multiple galaxies' spectra (\citealt{Cucciati+14}; Hayashino et al. in prep), which is also expected from numerical simulations \citep{Stark+15}. More generally, \citet{Mukae+16} found a weak correlation between the strength of the IGM H~{\sc i} absorption and the galaxy number density by analysing QSO spectra as background light sources and large number of photo-$z$ objects as foreground galaxies. \citet{Lee+14,Lee+16} established a novel approach to resolve the H~{\sc i} gas structure on Mpc scales by analysing the Ly$\alpha$ forest in individual spectra of background star-forming galaxies. They applied this three-dimensional H~{\sc i} reconstruction scheme (IGM H~{\sc i} tomography) to star-forming galaxies in the COSMOS field and found a correlation between galaxy onverdensity and H~{\sc i} absorption enhancement \citep{Lee+16}. 
The CGM structure has also been investigated by analysing the H~{\sc i} absorption as a function of the projected distance between the background sight-line object and the nearest foreground galaxy. An H~{\sc i} absorption enhancement was detected out to $\sim 2$\,physical Mpc from Lyman Break Galaxies (LBGs) at $z = 2$ -- $3$ \citep{Rakic+12,Rudie+12,Turner+14}. The CGM gas kinematics were also investigated by combining properties of the H~{\sc i} and metal absorption (e.g., \citealt{Turner+16}), which suggest the clumpy and outflowing nature \citep{Steidel+10,Crighton+15}. 

Investigating a relation between galaxies and H~{\sc i} gas in an extremely high density region is interesting in the context of both the large-scale structure formation and the environmental dependency of CGM properties. The SSA22 field, which we focus in this paper, contains one of the largest high density structures observed so far at high redshift. We previously performed a very wide-field imaging of $z = 3.1$ Ly$\alpha$ emitters (LAEs) with Subaru/Suprime-Cam \citep{Miyazaki+02} seven contiguous field-of-views (FoVs), showing that the prominently high density structure of LAEs is stretched over $\sim 50$\,comoving Mpc \citep{Yamada+12a}. Even at $> 50$\,comoving Mpc far from the LAE density peak, the LAE density averaged over a Suprime-Cam single FoV scale is larger than that estimated in control fields. Spectroscopy of the ``proto-cluster'' member galaxies revealed that the SSA22 large-scale structure seems to be a complex of filamentary structures and distinct clumps of galaxies in three-dimensional space \citep{Matsuda+05,Topping+16}. These geometrical properties suggest that the SSA22 porto-cluster is an ancestor of local superclusters or Great Wall \citep{Yamada+12a}. Significant number density excesses of photometrically selected Distant Red Galaxies (DRGs) and Submilimeter Galaxies (SMGs) were also reported \citep{Tamura+09,Uchimoto+12,Kubo+13,Umehata+14}, which were confirmed spectroscopically in the central small ($\la 20$\,arcmin$^2$) area \citep{Kubo+15,Umehata+15}. This suggest that assembly of massive galaxies already proceeded at least in the central region of the $z = 3.1$ porto-cluster. 

In the SSA22 field, we conducted a deep spectroscopic survey of LBGs using the VLT/VIMOS \citep{LeFevre+03}, and found a good positive-correlation between the LBG number density and the IGM H~{\sc i} absorption in a redshift range of $2.5 \la z \la 3.5$ (Hayashino et al. in prep). While a prominent H~{\sc i} absorption enhancement was confirmed at the proto-cluster redshift we could not resolve the H~{\sc i} absorption structure spatially, because we stacked background LBG spectra to increase a signal-to-noise ratio (S/N). While the majority of the previous studies for the intervening H~{\sc i} absorption are based on spectroscopy, we can identify the absorption photometrically by a fine-tuned narrow-band imaging in principle. If we focus on H~{\sc i} gas at $z = 3.1$, a filter with a central wavelength of $\sim 4980$\,\AA\ can be used to capture its Ly$\alpha$ absorption. A custom narrow-band filter $NB497$ has a central wavelength of $4977$\,\AA\ and a width of $78$\,\AA\ in Full Width Half Maximum (FWHM) \citep{Hayashino+04}, which limits the spatial resolution for H~{\sc i} gas structure along line-of-sight of $\sim 60$\,comoving Mpc in the $z = 3.1$ Universe. Such a coarse resolution is still enough to investigate the SSA22 large-scale structure because it is expected to spread out $\ga 50$\,comoving Mpc. 

In this paper, we introduce our new scheme to characterize the $z = 3.1$ H~{\sc i} absorption strength photometrically, and show the results of its application to the multi-band imaging data available in the SSA22 and two control fields. We briefly describe the data used in this study in Section~2, and introduce our method to measure the $z=3.1$ H~{\sc i} absorption strength in Section~3. We show the results in Section~4, and discuss them in Section~5. We use the AB magnitude system \citep{OkeGunn83} and adopt a cosmology with $H_{0}=70.4$ km s$^{-1}$ Mpc$^{-1}$, $\Omega_{M}=0.272$, and $\Omega_{\Lambda}=0.728$ \citep{Komatsu+11}.

\section{Data}

\subsection{Imaging data}\label{sec:data_image}

We collected multi-band imaging data available in the SSA22 field to measure the strength of Ly$\alpha$ absorption caused by the H~{\sc i} gas at $z = 3.09$. \citet{Yamada+12a} conducted $B$, $V$, and $NB497$ imaging observations with the Suprime-Cam equipped on the Subaru telescope. Among their wide survey coverage, 7 contiguous Suprime-Cam FoVs, we focus on the central FoV (SSA22-Sb1) where background sight-line galaxies at $z \ga 3.2$ are available from rich spectroscopic surveys (section~\ref{sec:data_bkggal}). There is the density peak of $z = 3.09$ LAEs in the SSA22-Sb1 field, and also the significant density excess of DRGs and SMGs was reported \citep{Tamura+09,Uchimoto+12,Kubo+13,Umehata+15}. We also used following images available in the SSA22-Sb1: Subaru/Suprime-Cam $R$, $i'$, and $z'$ band images \citep{Hayashino+04,Nakamura+11} and the UKIRT/WFCAM $K$ band image (UKIDSS DXS DR10; \citealt{Casali+07,Lawrence+07}). These images are smoothed so as to match the point spread function (PSF) size of bright stars to the FWHM of $1{\farcs}1$. For the smoothed images, we measured $5 \sigma$ limiting magnitudes with $2{\farcs}2$ diameter apertures, resulting in $B = 27.1$, $NB497 = 26.4$, $V = 26.8$, $R = 26.5$, $i' = 26.3$, $z' = 25.6$, and $K = 22.9$. We then constructed a muliti-band photometry catalog for $i'$ band detected objects. Object extraction was performed on the $i'$ band image using SExtractor \citep{BertinArnouts96} version 2.5.0. The $2{\farcs}2$ aperture magnitudes were measured in the all band images, where each aperture was centred on the $i'$ band object position. We corrected these magnitudes for the Galactic extinction with $A_B=0.246$, $A_{NB497}=0.215$, $A_V=0.190$, $A_R=0.154$, $A_i=0.124$, $A_z=0.090$, and $A_K=0.023$. These correction values were estimated for the centre of the SSA22-Sb1 ($\alpha = 22^{\rm h}17^{\rm m}33^{\rm s}, \delta=+00{\degr}15{\arcmin}08{\arcsec}$ in J2000.0) based on the work by \citet{Schlegel+98}, assuming $R_V=A_V/E(B-V)=3.1$. A spatial variation of $\Delta E(B-V) \sim 0.03$ in the SSA22-Sb1 area may make at most $\sim 0.1$\,mag systematic offsets in our estimates of the H~{\sc i} absorption strength ($\Delta NB497$; see the section~\ref{sec:dNB497_method}), however which is smaller than an excess of the $\Delta NB497$ measured in the SSA22-Sb1 field (section~\ref{sec:field2field_var}). 

For a comparison, we also constructed similar multi-band photometry catalogs in the Great Observatory Optical Deep Survey North (GOODS-N; \citealt{Dickinson+04}) field and the Subaru/$XMM-Newton$ Deep Survey (SXDS; \citealt{Furusawa+08}) field. In the GOODS-N field, we collected Subaru/Suprime-Cam $B$, $V$, and $NB497$ band images from \citet{Yamada+12a} and $R$, $i'$, and $z'$ band images from \citet{Capak+04}. These images are smoothed so as to match the PSF (FWHM $= 1{\farcs}2$), and their $5\,\sigma$ limiting magnitudes are $B = 26.8$, $NB497=26.7$, $V = 25.9$, $R = 26.3$, $i' = 25.7$, and $z' = 25.4$ with $2{\farcs}4$ diameter apertures. We constructed a multi-band photometry catalog for these images in the same manner as the SSA22-Sb1 catalog, where the objects were detected in the $V$ band image and $2{\farcs}4$ apertures were used for the photometry. The Galactic extinction corrections, $A_B=0.048$, $A_{NB497}=0.042$, $A_V=0.037$, $A_R=0.030$, $A_i=0.024$, and $A_z=0.018$, were applied, which are estimated at $(\alpha, \delta) = (12^{\rm h}37^{\rm m}22^{\rm s}, \delta=+62{\degr}11{\arcmin}33{\arcsec})$ in J2000 \citep{Schlegel+98}. 

In the SXDS field, Subaru/Suprime-Cam $B$, $V$, and $NB497$ images are available from \citet{Yamada+12a}. They observed three contiguous Suprime-Cam FoVs, SXDS-1C ($\alpha = 2^{\rm h}18^{\rm m}00^{\rm s}, \delta=-5{\degr}00{\arcmin}00{\arcsec}$), SXDS-2N ($\alpha = 2^{\rm h}18^{\rm m}00^{\rm s}, \delta=-4{\degr}34{\arcmin}59{\arcsec}$), and SXDS-3S ($\alpha = 2^{\rm h}18^{\rm m}00^{\rm s}, \delta=-5{\degr}25{\arcmin}01{\arcsec}$). \citet{Yamada+12a} matched their PSF sizes to FWHM $\approx 1{\farcs}0$ and measured $5\,\sigma$ limiting magnitudes as $B \approx 27.5$, $NB497 \approx 26.2$, and $V \approx 27.1$ with $2{\farcs}2$ diameter apertures. We performed object extraction on the $V$ band image and measured $2{\farcs}0$ diameter aperture magnitudes in the $B$, $V$, and $NB497$ images. For the $R$, $i'$, and $z'$ band photometry, we used a public catalog \citep{Furusawa+08}. Their $5\,\sigma$ limiting magnitudes are $R \approx 27.0$, $i' \approx 26.9$, and $z' \approx 25.8$ with $2{\farcs}0$ diameter apertures. We merged our own $V$ band detection catalog and the public catalog using a matching radius of $1{\farcs}0$. Here, $V$ band photometry included in the public catalog were used to correct the photometry scheme difference between \citet{Furusawa+08} and ours: we corrected the $R$, $i'$, and $z'$ band magnitudes by adding the public catalog colours ($V-R$, $V-i'$, and $V-z'$) to the $V$ band magnitudes that we measured. All the magnitudes were corrected for the Galactic extinction \citep{Schlegel+98}: $A_B=0.085/0.077/0.093$, $A_{NB497}=0.075/0.067/0.082$, $A_V=0.066/0.059/0.072$, $A_R=0.056/0.052/0.062$, $A_i=0.044/0.040/0.048$, and $A_z=0.031/0.029/0.034$ for the SXDS-1C/2N/3S field, respectively.

\subsection{Background light sources at $z \approx 3.4$}\label{sec:data_bkggal}

Ly$\alpha$ absorption by the $z = 3.1$ H~{\sc i} gas is imprinted in the spectrum of every galaxy behind the gas at wavelength $\lambda \approx 4980$\,\AA\ in the observer's rest-frame (hereafter, the observer-frame). This is in the $NB497$ filter wavelength coverage. In this work, we used only galaxies with a spectroscopic redshift of $3.29 \leq z \leq 3.54$ as background light sources to isolate the $z = 3.1$ H~{\sc i} Ly$\alpha$ absorption from interstellar absorption lines of the background galaxies themselves. Wavelengths of the $NB497$ filter ($4937.5 - 5015.7$\,\AA\ in the observer-frame) shift to a portion of $\lambda = 1087 - 1170$\,\AA\ in the $z = 3.29 - 3.54$ galaxy's rest-frame, where N~{\sc ii}\,$\lambda 1084$ and C~{\sc iii}\,$\lambda 1178$ as well as broad wings of Ly$\alpha$ and Ly$\beta$ of the galaxies themselves do not contaminate.  

We gathered galaxies at $z = 3.29$ -- $3.54$ from the following spectroscopic redshift catalogs. In the SSA22-Sb1 field, we used the catalogs from \citet{Steidel+03}, \citet{Nestor+13}, \citet{Saez+15}, and our own observations using the VLT/VIMOS and Keck/DEIMOS \citep{Faber+03}: VIMOS06 \citep{Kousai+11}, VIMOS08 (\citealt{Kousai+11}; Hayashino et al. in prep), VIMOS12 (Umehata et al. in prep; Mawatari et al. in prep), DEIMOS08 (Otsuka et al. in prep), and DEIMOS15 (Mawatari et al. in prep). In the GOODS-N field, we used the 3D-HST catalog\footnote{http://3dhst.research.yale.edu/Data.php} \citep{Momcheva+16}. In the SXDS field, we used the catalogs from the UDSz project\footnote{http://www.nottingham.ac.uk/astronomy/UDS/UDSz/} \citep{Bradshaw+13,McLure+13}, the 3D-HST \citep{Momcheva+16}, and the project led by Akiyama, Simpson, Croom, Geach, Smail and van Breukelen\footnote{http://www.nottingham.ac.uk/astronomy/UDS/data/data.html} (hereafter, we call the UDS-AGN project; \citealt{Smail+08,Simpson+12,Akiyama+15}). Since redshifts in the 3D-HST catalog which were determined from the combination of the $HST$ grism spectra and multi-band photometry have relatively large uncertainty we forced them to satisfy $z_{\rm min} \ge 3.29$ and $z_{\rm max} \le 3.54$, where $z_{\rm min}$ and $z_{\rm max}$ are the lower and upper 68\,\% confidence limits for the grism redshifts \citep{Momcheva+16}. We collected 77, 96, and 80 background galaxies at $z = 3.29 - 3.54$ in the SSA22-Sb1, GOODS-N, and SXDS fields, respectively. The colours of the background galaxies in the SSA22-Sb1 field are systematically bluer than those in the two control fields, which is due to the difference in the spectroscopic target selections. In the SSA22-Sb1 field, almost all surveys targeted the UV-selected LBGs and narrow-band selected LAEs which generally have blue UV slope. In the other two fields, however, majority of the background galaxies were collected from the 3D-HST survey which is based on the near-infrared ($HST$/WFC3) slitless spectroscopy and can detect more dusty red galaxies.

\subsection{Galaxies at $z = 3.1$}

We gathered $z = 3.1$ galaxies from literature to investigate the relation between H~{\sc i} gas and galaxies in the $z = 3.1$ Universe. We first collected photometrically selected LAEs. \citet{Yamada+12a} used the $NB497$ filter to construct a homogeneous sample of LAEs with the rest-frame equivalent width of ${\rm EW}_0 \ga 50$\,\AA\ at $z = 3.06$ -- $3.13$ in the three our target fields. We also collected spectroscopically confirmed galaxies at $z = 3.06$ -- $3.13$. In the SSA22-Sb1 field, we used the spectroscopic redshift catalogs from \citet{Steidel+03}, \citet{Matsuda+05}, \citet{Matsuda+06}, \citet{Yamada+12b}, \citet{Nestor+13}, \citet{Saez+15}, and our own observations: VIMOS06, VIMOS08, VIMOS12, DEIMOS08, DEIMOS15, and the Magellan/IMAX \citep{Dressler+11} observation (P.I.: M. Ouchi; hereafter, oIMACS). In the GOODS-N field, we used the 3D-HST catalog \citep{Momcheva+16}. In the SXDS field, we used the catalogs from the UDSz project \citep{Bradshaw+13,McLure+13}, the 3D-HST \citep{Momcheva+16}, and the UDS-AGN project \citep{Smail+08,Simpson+12,Akiyama+15}. For the 3D-HST catalog objects, we applied the same manner as adopted in the selection of the background galaxies (section~\ref{sec:data_bkggal}). These spectroscopically confirmed $z = 3.1$ galaxy samples contain some of the photometrically selected LAEs.

\section{Analysis}

\subsection{Careful calibration of photometric colours} \label{sec:color_calib}

\begin{figure}
\begin{center}
\includegraphics[width=1.0\linewidth, angle=0]{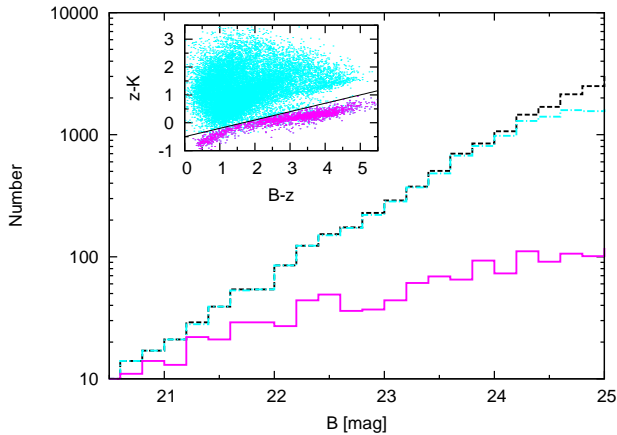}
\caption{Number counts of all objects detected in the $i'$ band with a $3 \sigma$ significance (black dashed histogram), objects detected in all of $i$, $B$, $z$, and $K$ bands with a $3 \sigma$ significance (cyan dot-dashed histogram), and stars selected in the $z-K$ versus $B-z$ two colour diagram (magenta solid histogram) in the SSA22-Sb1 field. Sub-panel shows objects detected in $i$, $B$, $z$, and $K$ bands (cyan points) and stars (magenta points) in the $z-K$ versus $B-z$ two colour diagram, where black line corresponds to the boundary between galaxies and stars \citep{Daddi+04}. \label{fig:BzKstar}}
\end{center}
\end{figure}

\begin{figure*}
\begin{center}
\includegraphics[width=1.0\linewidth, angle=0]{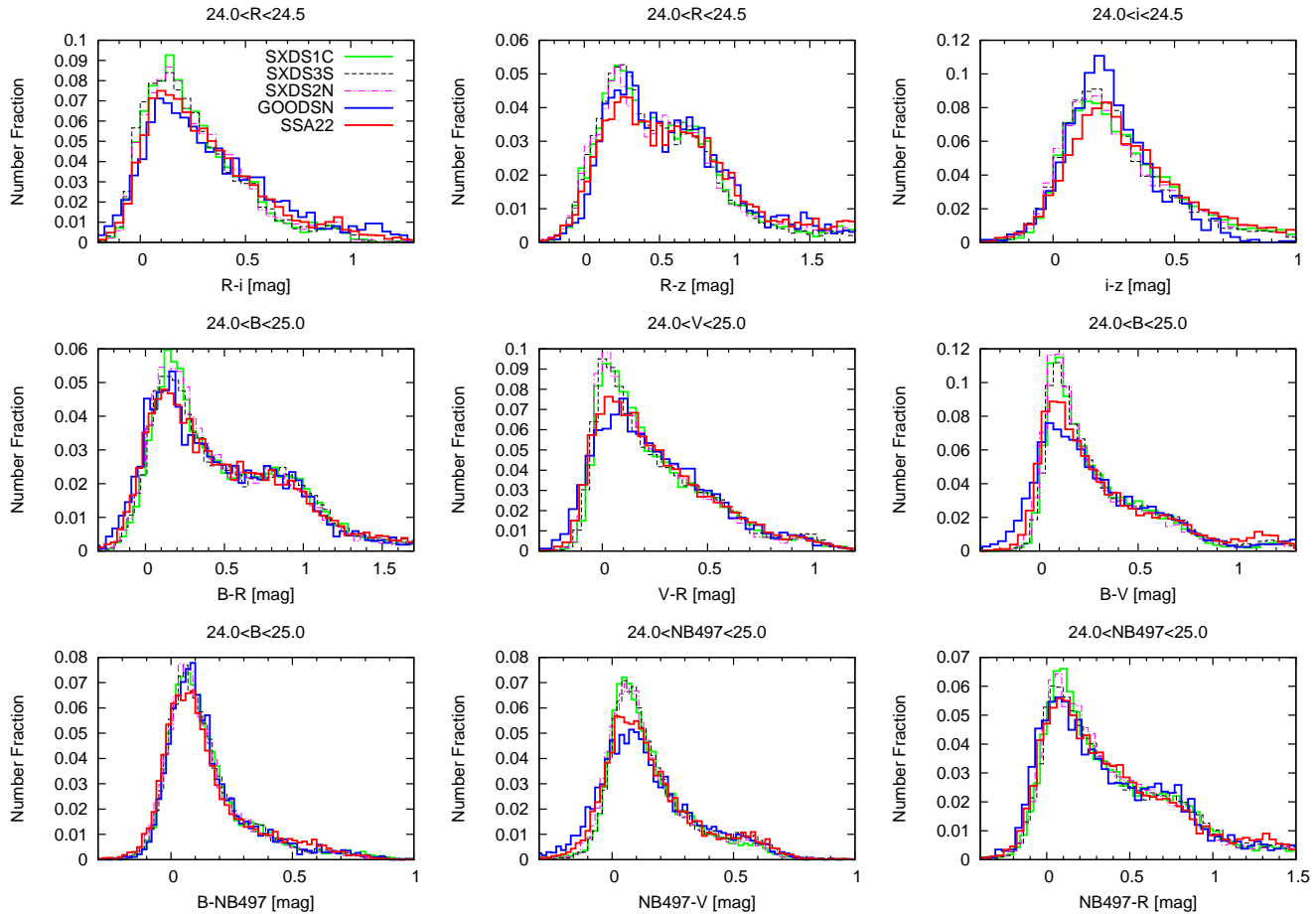}
\caption{Various colour distributions in the SSA22-Sb1 (red solid histogram), SXDS-1C (green solid histogram, reference), SXDS-3S (black dashed histogram), SXDS-2N (magenta dot-dashed histogram), and GOODS-N (blue solid histogram) fields. The magnitude range of the objects used are shown in the top of each panel. We adopted corrections to the $B$, $V$, $R$, $i$, and $NB497$ magnitudes so that all colour distributions become consistent among the fields. \label{fig:zerocorr}}
\end{center}
\end{figure*}

In order to characterize the $z = 3.1$ H~{\sc i} absorption strength photometrically, we need to calibrate photometric measurements carefully. There may be $\ga 0.1$\,mag systematic uncertainties on magnitudes and colors caused by the following two effects. Our photometry using the fixed apertures is very sensitive to the image smoothing which was adopted so that the PSF sizes are matched to each other (section~\ref{sec:data_image}); there might be systematic offsets in the photometric zero-points among fields because the imaging data used in this work are based on the several observations and reductions by independent research groups. In this section, we adopt corrections for band photometry so that the colour distribution in every field is matched to that in the SXDS-1C field. 


Using extragalactic objects is desirable to investigate any difference in colour distributions among the target fields, because it is difficult to know the accurate amount of the Galactic dust extinction for each galactic star without its three-dimensional location. We investigated the fraction of stars in all objects in the SSA22-Sb1 field, where stars were selected in the $B-z$ versus $z-K$ two colour diagram by the same manner as \citet{Daddi+04}. The two colour diagram and the number counts of the selected stars and all objects are shown in Figure~\ref{fig:BzKstar}, which shows that at $B > 24$\,mag the fraction of stars is at most $\sim 10$\,\% and the stars are not likely to change colour distributions significantly. The stellar fraction among all objects should be smaller in the GOODS-N and SXDS fields because these control fields are oriented towards outer parts of the Milky Way compared with the direction of the SSA22-Sb1 field. From these, we neglect the stellar contribution to colour distributions of faint objects with $\ga 24$\,mag in all the fields. 

Then, we compared various colours corrected for the Galactic extinctions among the SSA22-Sb1, GOODS-N, SXDS-1C, SXDS-2N, and SXDS-3S fields. We carefully selected the magnitude range of objects used for each colour distribution; fainter than $24$\,mag for rejection of stars and much brighter than the $5\sigma$ limiting magnitude for the sampling completeness. We estimated correction magnitudes for the photometric zero-points of the all band images in the all fields to match the $R-i$, $i-z$, $R-z$, $B-R$, $V-R$, $B-V$, $NB497 - R$, $B - NB497$, and $NB497 - V$ colour distributions to those in the SXDS-1C field. The correction values are typically $< 0.1$\,mag, and $0.16$\,mag at maximum. The colour distributions after adopting the photometric zero-point corrections are shown in Figure~\ref{fig:zerocorr}. 

Since the colour differences in the all fields relative to the reference field (the SXDS-1C field) are minimized, relative photometric uncertainties among the fields are expected to be at most $0.04$\,mag (corresponding to two bins shifts in Figure~\ref{fig:zerocorr}). On the other hand, we here mention that the set of the correction magnitudes adopted in this work is not a unique solution, and there may be other sets yielding a good match in the colour distributions among the fields. Furthermore, the reference colour distribution in the SXDS-1C field may be slightly different from the actual distributions. In the absolute sense, typically $\sim 0.1$\,mag systematic uncertainty still remains in each band photometry in each field. For example, if we observe galaxies with the same intrinsic SEDs in every field, the observed SEDs are possibly different from the intrinsic one by $\sim 0.1$\,mag but the difference of the observed SEDs relative to that in the reference field should be $\la 0.04$\,mag. These absolute and relative uncertainties in the photometry in the observed fields may cause systematic errors in our estimates of the H~{\sc i} absorption strength ($\Delta NB497$; see the following section).

\subsection{$\Delta$NB497 as a tracer of $z=3.1$ H~{\sc i} absorption} \label{sec:dNB497_method}

\begin{figure}
\begin{center}
\includegraphics[width=1.0\linewidth, angle=0]{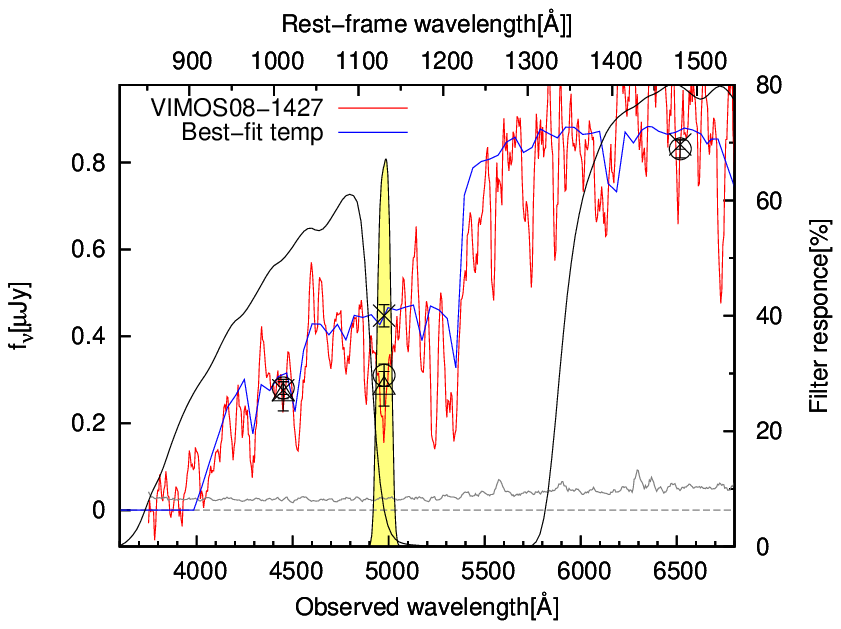}
\caption{An example of our method to characterize the $z = 3.1$ H~{\sc i} absorption. The observed $B$, $NB497$, and $R$ magnitudes are shown by black circles, where each filter response curve is shown by black curves. The blue spectrum is the best-fit template for the $B$, $R$, $i$, and $z$ band photometry. The observed spectrum and noise for this object are shown by red and grey curves, which were obtained in the VIMOS08 observations. Black crosses and triangles are the filter convolved magnitudes for the best-fit template and observed spectrum. $R$ band filter convolved magnitude for the observed spectrum is not shown because the spectral coverage does not include the whole $R$ band wavelengths. \label{fig:method_idea}}
\end{center}
\end{figure}

\begin{table}
\caption{Summary of background light source numbers.\label{tb:Nsummary}}
\begin{tabular}{lccc}
\hline \hline
Field & N$_{\rm bkg}$ $^{\rm a}$ & N$_{\rm bright}$ $^{\rm b}$ & N$_{\rm use}$ $^{\rm c}$ \\
\hline
SSA22-Sb1 & 77 & 74 & 60\\
GOODS-N & 83 & 47 & 40\\
SXDS & 70 & 53 & 33\\
\hline
\end{tabular}
\\
\noindent
$^{\rm a}${Number of galaxies at a spectroscopic redshift $3.29 \leq z \leq 3.54$ as background light sources for $z = 3.1$ H~{\sc i} absorption.}\\
$^{\rm b}${Number of the background galaxies which are bright enough to be detected in the $B$, $NB497$, $R$, $i$, and $z$ band images.}\\
$^{\rm c}${Number of the background galaxies used for our discussion, whose SEDs are well fit by the \citet{BruzualCharlot03} model templates with a reduced $\chi^2$ value less than or equal to $2$.}
\end{table}

\begin{figure}
\begin{center}
\includegraphics[width=1.0\linewidth, angle=0]{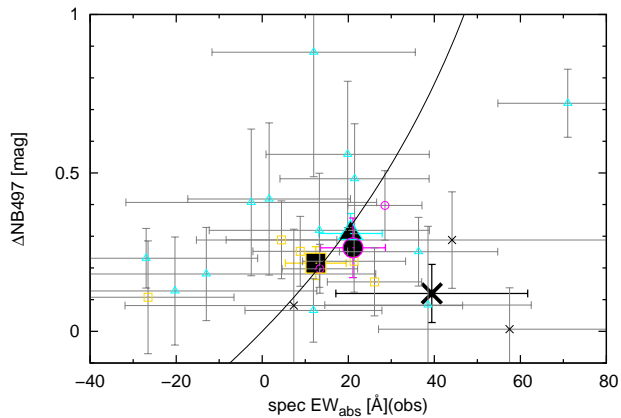}
\caption{Comparison of the $\Delta NB497$ estimated from the multi-band photometry and the absorption EW in the observer-frame measured from the spectroscopic data. We used the spectra of 23 galaxies at $z = 3.29 - 3.54$, which were obtained in the VIMOS06 and VIMOS08 observations. Symbol types are divided by the S/N in the DA region of each spectrum (S/N $< 3$: black crosses; $3 <$ S/N $< 7$: cyan triangles; $7 <$ S/N $< 10$: yellow squares; S/N $> 10$: magenta circles), where big filled symbols correspond to the weighted mean values. Black solid curve shows the expected relation: $\Delta NB497 = -2.5 \log (1 - {\rm EW}/78$\AA$)$. Red cross shows the weighted mean value of the 23 galaxies used. \label{fig:dM2specEW}}
\end{center}
\end{figure}

In this section, we introduce our scheme to characterize Ly$\alpha$ absorption by $z = 3.1$ H~{\sc i} gas photometrically. Among the background galaxies at $z = 3.24$ -- $3.59$ in the SSA22-Sb1, GOODS-N, and SXDS fields, we here focus on the 74, 57, and 61 objects bright enough to be detected in the $R$, $i$ (with more than $3\sigma$ significance), $B$, $z$ (with more than $2\sigma$ significance), and $NB497$ (with more than $1\sigma$ significance) band images. 

We performed a Spectral Energy Distribution (SED) fitting for $B$, $R$, $i$, and $z$ magnitudes of each background galaxy with a public code, Hyperz \citep{Bolzonella+00}. The $V$ band photometry was not used because the Ly$\alpha$ line of the background galaxies themselves contributes to the $V$ band flux, which is hard to be expected due to scattering and absorption by surrounding H~{\sc i} gas and dust. Spectral templates consist of the stellar synthesis models of \citet{BruzualCharlot03} with an exponentially declining star formation rate (SFR; $\tau = 1$\,Gyr) and a fixed metallicity ($Z = 0.004$). The IMF was fixed to a Chabrier IMF \citep{Chabrier+03} with the mass range of $0.1 M_{\odot}$ -- $100 M_{\odot}$. The template redshift was fixed to the spectroscopic redshift of each background galaxy. Model ages are forced to be less than the age of the Universe at the galaxies' redshift. Dust attenuation $A_V$ was changed from $0$ to $3$, where the \citet{Calzetti+00} dust attenuation law was adopted. In the Hyperz code, the template flux at a wavelength shorter than Ly$\alpha$ is attenuated by a mean IGM absorption at a given redshift according to \citet{Madau+95}. We have confirmed that the mean IGM attenuation at a wavelength range between the Ly$\alpha$ and the Ly$\beta$ (so called DA range) is consistent with a more recent model \citep{Inoue+14}. We obtained the best-fit template for each background galaxy, from which the $NB497$ magnitude was estimated ($NB497_{\rm temp}$). Uncertainty in the $NB497_{\rm temp}$ was estimated from the $1\sigma$ confidence interval ($\Delta \chi^2 \le 2.3$) in the fitting parameter space ($A_V$ versus model age). We measured the magnitude offset between the observed and template $NB497$ magnitudes ($\Delta NB497 = NB497_{\rm obs} - NB497_{\rm temp}$), which is expected to reflect strength of the $z = 3.1$ H~{\sc i} Ly$\alpha$ absorption relative to the average in the $z = 3.1$ Universe. An example of our method is shown in Figure~\ref{fig:method_idea}. In the following discussions, we use $\Delta NB497$ estimates of the background galaxies whose SED are well fit by a template with a reduced $\chi^2$ ($\chi_{\nu}^2$) $< 2$. In other words, we effectively remove AGNs, QSOs, and other exotic types of galaxies whose SEDs are very different from the star-forming model templates from our sample. The numbers of the background galaxies that we finally used are 60, 47, and 40 in the SSA22-Sb1, GOODS-N, and SXDS fields, respectively. Table~\ref{tb:Nsummary} is a summary of the numbers of the galaxies. 

As mentioned in the section~\ref{sec:color_calib}, the $\Delta NB497$ may have systematic errors which are propagated from those in the individual band photometry. We here assume that the possible systematic errors associated with the $\Delta NB497$ is $\sim 0.1$\,mag in the absolute sense and $\sim 0.04$\,mag in a relative sense among the three fields. While in the following we do not include these estimates of systematic errors in the $\Delta NB497$ measurements, we should always be aware of these possible systematic uncertainties. 

We checked the reliability of our $\Delta NB497$ scheme using spectra of some background galaxies in the SSA22-Sb1 fields. Our previous VLT/VIMOS observations (VIMOS06 and VIMOS08; \citealt{Kousai+11}; Hayashino et al. in prep) allows us to investigate H~{\sc i} Ly$\alpha$ absorption for individual spectra of some background galaxies. We measured the EWs of the $z = 3.1$ H~{\sc i} Ly$\alpha$ absorption in the available VIMOS spectra, where the absorption flux was estimated in the $NB497$ wavelength range and the continuum was estimated by averaging flux in the DA range except for the $NB497$ coverage in the observer-frame. The relation between $\Delta NB497$ and observer-frame EW can be described in an analytical form as $\Delta NB497 = -2.5 \log (1 - {\rm EW}/78$\AA$)$, where we assume that only $z = 3.1$ H~{\sc i} absorption deviates from the cosmic average in the redshift range corresponding to the $NB497$ filter coverage (FWHM $= 78$\,\AA). Figure~\ref{fig:dM2specEW} shows the comparison between the $z = 3.1$ H~{\sc i} absorption EWs measured spectroscopically and their $\Delta NB497$. In spite of the large uncertainties especially in the spectroscopic EWs, a better correlation can be seen for the background galaxies whose spectra have a higher S/N ($\ga 10$). We confirmed that the weighted mean values for subsamples divided by the spectral S/N are consistent with the expected relation except for the lowest S/N subsample. These suggests that we reasonably trace the absorption strength with the $NB497$ photometry and the continuum level with the best-fit model templates as shown in Figure~\ref{fig:method_idea}. 

The background sight-line galaxies in the SSA22-Sb1 field have bluer SEDs than those in the two control fields as mentioned in the section~\ref{sec:data_bkggal}, which possibly causes a bias in the $\Delta NB497$ measurements. We constructed a $z \approx 3.4$ mock galaxy sample to investigate the dependency of the $\Delta NB497$ on the background galaxy SEDs. The mock galaxy SEDs were generated from the redshifted \citet{BruzualCharlot03} model templates with different $R - z$ colours. Each model template spectrum was normalized so that the $R$ band magnitude becomes the typical value of the observed background galaxies ($R = 25$\,mag), and the each band magnitude was randomly extracted from a Gaussian probability distribution with the normalized model magnitude as the mean and the typical photometric uncertainty ($\sim 0.1$\,mag) as the standard deviation. We performed the SED fitting for the mock galaxies and estimated the $\Delta NB497$ with the same manner as described above. The resultant $\Delta NB497$ values distribute around zero and show no dependency on the input SED colours, which supports that our $\Delta NB497$ scheme properly evaluates the strength of $z = 3.1$ H~{\sc i} absorption relative to the cosmic average.

\section{Result}

\subsection{H~{\sc i} absorption enhancement over the entire SSA22-Sb1 field}\label{sec:field2field_var}

\begin{figure}
\begin{center}
\includegraphics[width=1.0\linewidth, angle=0]{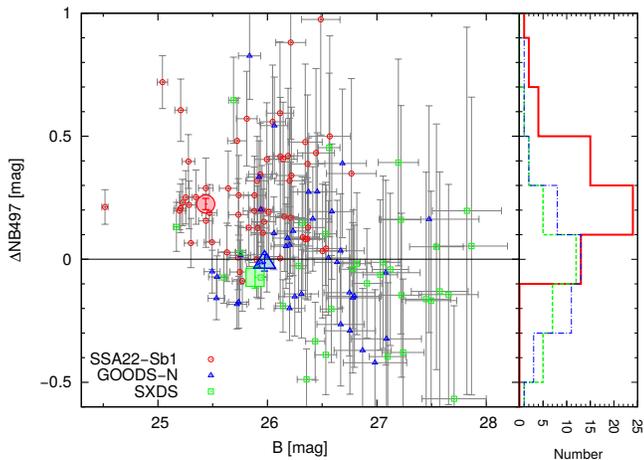}
\caption{Left panel shows the $\Delta NB497$ values as a function of the $B$ band magnitudes in the SSA22-Sb1 (red open circles), GOODS-N (blue open triangles), and SXDS (green open squares) fields. Big filled symbols correspond to the weighted mean value in each field. Right panel shows the number histogram of the $\Delta NB497$ (SSA22-Sb1: red solid histogram; SXDS: green dashed histogram; GOODS-N: blue dot-dashed histogram). \label{fig:Fcompari_dM}}
\end{center}
\end{figure}

\begin{figure}
\begin{center}
\includegraphics[width=1.0\linewidth, angle=0]{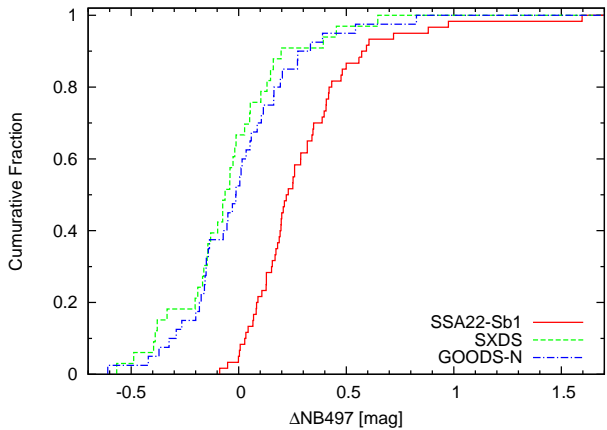}
\caption{Cumulative fraction histogram for the $\Delta NB497$ values in the three fields. \label{fig:Fcompari_dMhist}}
\end{center}
\end{figure}

First, we investigate field-to-field variance in the  $\Delta NB497$ measurements. Figure~\ref{fig:Fcompari_dM} shows the $\Delta NB497$ values as a function of the $B$ band magnitudes of the background sight-line galaxies. The $\Delta NB497$ values in the SSA22-Sb1 field are systematically larger than those in the GOODS-N and SXDS fields, which suggests an H~{\sc i} gas excess in the SSA22-Sb1 field. Weighted mean $\Delta NB497$ values, where  we used the inverse of the uncertainty as the weight, are $0.22 \pm 0.02$, $-0.01 \pm 0.03$, and $-0.07 \pm 0.04$ in the SSA22-Sb1, GOODS-N, and SXDS fields, respectively, where we adopt larger one between the internal and external errors as the uncertainty. Even after taking account of the possible relative systematic uncertainties of $\sim 0.04$\,mag among the three fields (see the section~\ref{sec:dNB497_method}), the excess of the $\Delta NB497$ in the SSA22-Sb1 field compared to the other fields is still significant. The negative $\Delta NB497$ values in the SXDS field is less significant, and we cannot claim from only the current data that the H~{\sc i} absorption is suppressed in the SXDS field relative to the cosmic average. The background sight-line galaxies in the GOODS-N and SXDS fields are systematically fainter in the $B$ band, which causes larger $\Delta NB497$ uncertainties but is less likely to cause the $\Delta NB497$ difference among the fields. Figure~\ref{fig:Fcompari_dMhist} shows cumulative histograms for the $\Delta NB497$ values in the three fields. We found a clear difference of the SSA22-Sb1 from the other two fields. We applied a Kolmogorov-Smirnov (K-S) test to check the difference of the $\Delta NB497$ values. Resultant K-S probabilities are $6.7 \times 10^{-7}$ in the SSA22-Sb1 versus GOODS-N, $2.6 \times 10^{-8}$ in the SSA22-Sb1 versus SXDS, and $0.69$ in the GOODS-N versus SXDS. A null hypothesis that the $\Delta NB497$ values in the SSA22-Sb1 field are extracted from the same mother sample as those in the other two fields is significantly rejected, while the $\Delta NB497$ distributions in the GOODS-N and SXDS fields are very consistent. 

\begin{figure*}
\begin{center}
\includegraphics[width=1.0\linewidth, angle=0]{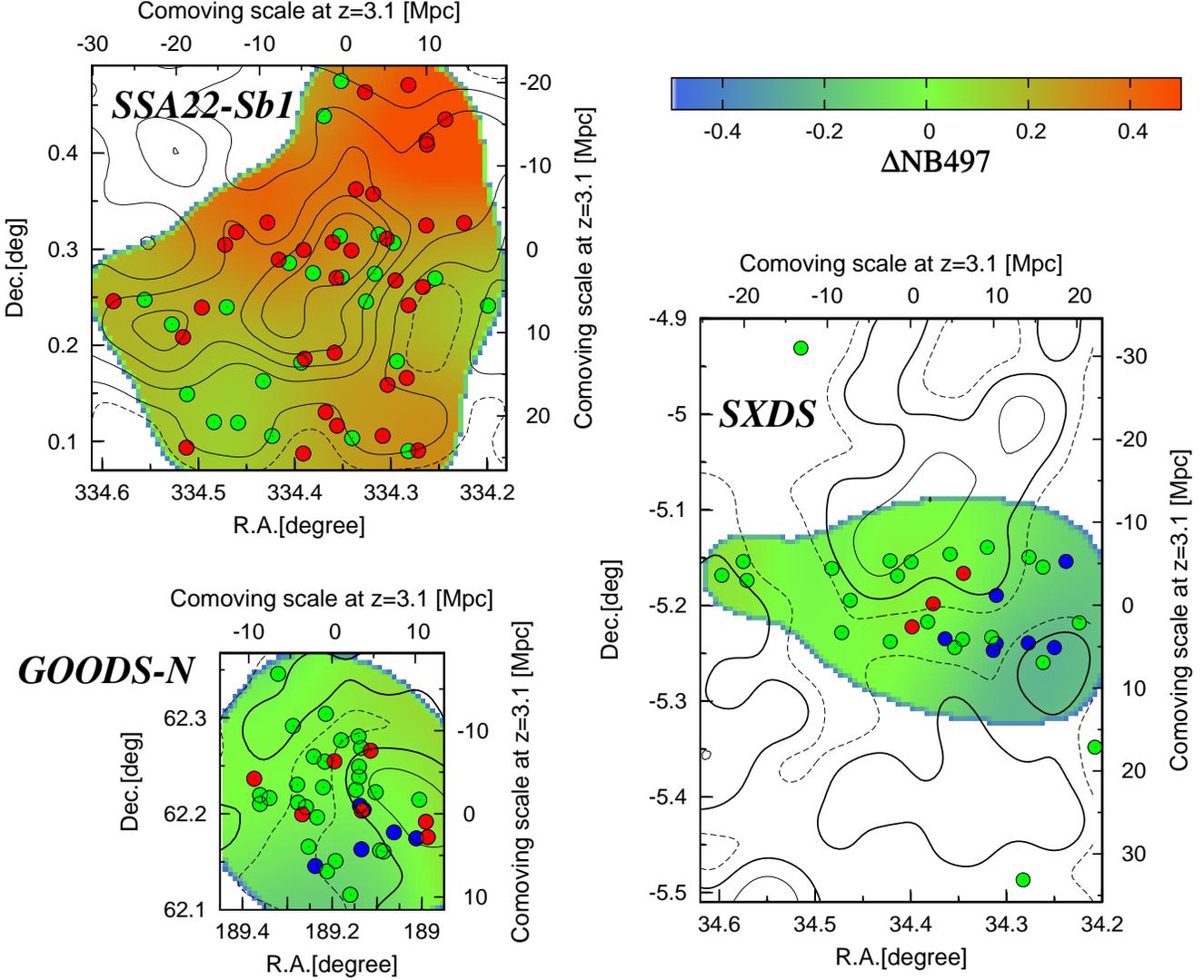}
\caption{Sky distribution of the background sight-line galaxies in the SSA22-Sb1, GOODS-N, and SXDS fields are shown by circles, where the symbol colours indicate the $\Delta NB497$ values ($\Delta NB497 > 0.2$: red; $-0.2 < \Delta NB497 < 0.2$: green; $\Delta NB497 <-0.2$: blue). Colour map shows the $\Delta NB497$ distribution; the $\Delta NB497$ values for every background sight-line galaxy were smoothed by a Gaussian kernel with $\sigma = 3'$ and divided by the smoothed sight-line number density. Black contours show the smoothed and normalized number density of the photometric LAEs: $0.5 \times$ (dashed), $1 \times$ (thick), $2 \times$, $3 \times$, $4 \times$, and $5 \times$ the average density ($0.2$\,arcmin$^{-2}$; \citealt{Yamada+12a}). \label{fig:dM_skymap_allF}}
\end{center}
\end{figure*}

Since we used faint star-forming galaxies as the background light sources, we achieved a high spatial resolution of sight-lines for $z = 3.1$ H~{\sc i} absorption: mean separations of the background sight-line galaxies are $\sim 2{\arcmin}$ -- $3{\arcmin}$ or $\sim 4.1$ -- $5.5$ comoving Mpc at $z = 3.1$. We visualized two-dimensional sky maps of the $z = 3.1$ H~{\sc i} absorption strength, where each $\Delta NB497$ is smoothed by a Gaussian kernel with $\sigma = 3{\arcmin}$ and divided by the smoothed sight-line number density. Figures~\ref{fig:dM_skymap_allF} show the two-dimensional smoothed $\Delta NB497$ maps as well as sky distributions of the background sight-line galaxies. We only show the smoothed $\Delta NB497$ maps where the smoothed number density of background sight-line galaxies is more than $150$\,degree$^{-2}$. In Figures~\ref{fig:dM_skymap_allF}, contour lines show the local number density of photometric LAEs \citep{Yamada+12a} obtained after being smoothed by a Gaussian kernel with $\sigma = 1{\arcmin}.5$. We can see the large scale $\Delta NB497$ excess in the SSA22-Sb1 field, which extends to $\sim 50$\,comoving Mpc in the $z = 3.1$ Universe. On the other hand, no clear spatial alignment between the smoothed $\Delta NB497$ and $z = 3.1$ LAE number density can be seen at least by eyes in all the fields.

\subsection{$\Delta NB497$ as a function of environment}

\begin{figure}
\begin{center}
\includegraphics[width=1.0\linewidth, angle=0]{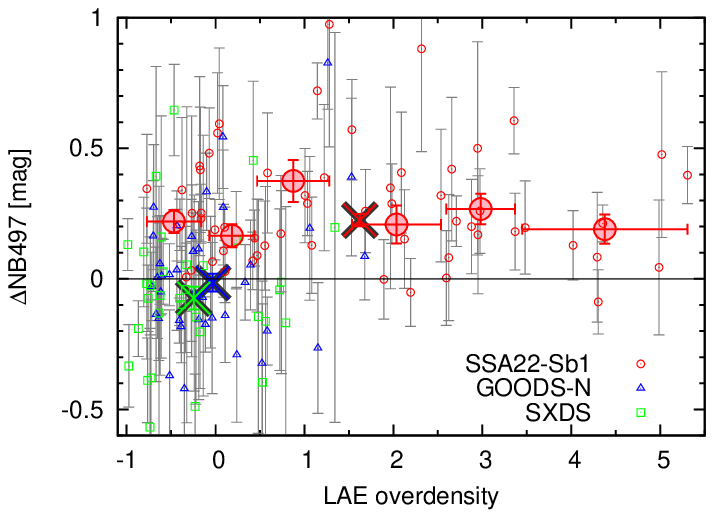}
\caption{$\Delta NB497$ as a function of the LAE overdensity. Red circles, blue triangles, and green squares show the sight-lines in the SSA22-Sb1, GOODS-N, and SXDS fields, respectively. Big crosses correspond to the weighted mean value in each field. Big filled circles are the weighted mean values for every 10 data points along the LAE overdensity in the SSA22-Sb1 field. \label{fig:dM2ovden}}
\end{center}
\end{figure}

\begin{figure}
\begin{center}
\includegraphics[width=1.0\linewidth, angle=0]{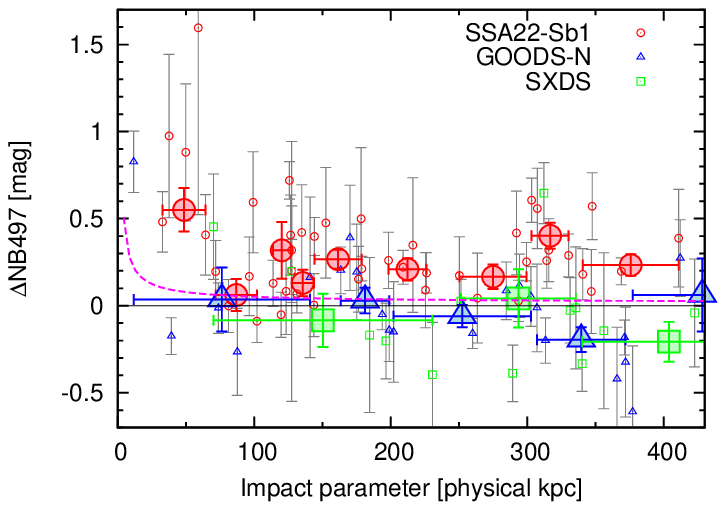}
\caption{$\Delta NB497$ as a function of distance from the nearest $z = 3.1$ object. Red circles, blue triangles, and green squares show the sight-lines in the SSA22-Sb1, GOODS-N, and SXDS field, respectively. Big filled symbols correspond to the weighted mean values for every five data points in each field. The magenta dashed curve shows the $b$ -- $\Delta NB497$ relation estimated from the H~{\sc i} absorption EWs of \citet{Rakic+12}, where we assumed the simple analytic form as $\Delta NB497 = -2.5 \log \{1 - {\rm EW^{abs}_{rest}}\times(1+3.1)/78$\AA$\}$. \label{fig:dM2galdist}}
\end{center}
\end{figure}

In this section, we investigate strength of the $z = 3.1$ H~{\sc i} absorption, $\Delta NB497$, as a function of environment around each sight-line. Photometric LAEs are considered to be trace environment although it may deviate from the real shape of underlying dark matter structure due to a small duty cycle of LAEs ($\sim 1$\,\%; \citealt{Ouchi+10}). We estimated the LAE overdensity, $\delta_{LAE} = (n_{LAE}-\overline{n})/\overline{n}$, at the position of each background sight-line, where we adopted the average LAE number density in general fields reported by \citet{Yamada+12a} of $\overline{n} = 0.2$\,arcmin$^{-2}$ and the LAE local number density ($n_{LAE}$) smoothed with a $\sigma = 1{\arcmin}.5$ gaussian kernel. Figure~\ref{fig:dM2ovden} shows the $\Delta NB497$ values as a function of $\delta_{LAE}$ for the background sight-lines in all the fields. SSA22-Sb1 field contains high density regions with $\delta_{LAE} \ga 2$ compared with the other two control fields. The weighted mean values for the three fields (big crosses in Figure~\ref{fig:dM2ovden}) suggest a trend that the average $\Delta NB497$ becomes higher with a larger average LAE overdensity. The negative value of the $\Delta NB497$ in the SXDS field may be explained by the negative LAE overdensity, as well as by possible systematic errors (see the section~\ref{sec:field2field_var}). However, such a trend between the LAE overdensity and the $\Delta NB497$ cannot be found if we look at only the data of the SSA22-Sb1 field. We took every 10 data points in the SSA22-Sb1 field along the LAE overdensity and calculated their weighted mean $\Delta NB497$ values (filled circles in Figure~\ref{fig:dM2ovden}). The binned $\Delta NB497$ averages flatly distribute along the LAE overdensity, and even at the smallest $\delta_{LAE}$ bin, the $\Delta NB497$ value in the SSA22-Sb1 field is significantly larger than those in the other two control fields. 

The resolution scale on environment traced by LAEs is roughly equal to the mean separation of the LAEs, which is $\sim 3$ comoving Mpc in the $z = 3.1$ Universe. The strength of H~{\sc i} absorption may be more sensitive to a smaller scale: H~{\sc i} gas abundance in the CGM. We selected the nearest galaxy at $z = 3.1$ from each background sight-line and measured their projected distance (impact parameter: $b$). Figure~\ref{fig:dM2galdist} shows the $\Delta NB497$ values as a function of the impact parameter. We also took every five data points in each field along the impact parameter and calculated their weighted mean $\Delta NB497$ (filled symbols in Figure~\ref{fig:dM2galdist}). We only look at the distance less than $430$\,physical kpc, which is an average impact parameter for star-forming galaxies with $M_{UV} < -20$ at $3.06 < z < 3.13$ expected from UV luminosity functions of \citet{ReddySteidel09} and \citet{SawickiThompson06}. There may be nearer but unobserved galaxies at $z = 3.1$ for the background sight-lines with large impact parameters of $b > 430$\,physical kpc. In Figure~\ref{fig:dM2galdist}, the $\Delta NB497$ seems to be enhanced at the smallest impact parameter ($b < 100$\,physical kpc) at least in the SSA22-Sb1 field, which can be interpreted as the effect of H~{\sc i} gas halos associated with these nearest galaxies. In the control fields, such a $\Delta NB497$ enhancement at $b < 100$\,physical kpc is unclear due to a small number of the data points. At a larger impact parameter, the $\Delta NB497$ values in the SSA22-Sb1 field seem to be constant and significantly above zero, which is contrast to the $\Delta NB497$ distribution in the two control fields. 


\section{Discussion}

\subsection{What drives the large-scale difference in the H~{\sc i} absorption?}\label{sec:HI_LSS}

We obtained the notable result that the Ly$\alpha$ absorption by the H~{\sc i} gas at $z = 3.1$ over the entire SSA22-Sb1 field is stronger than that in the control fields. Interestingly, the $\Delta NB497$ estimates are not correlated with the local ($\sim 3$\,comoving Mpc scale) LAE overdensity or the impact parameter if $> 100$\,physical kpc. A remarkable difference between the SSA22 and the two control fields is that the former field contains the prominent density excess of various types of galaxies \citep{Steidel+98,Hayashino+04,Uchimoto+12,Kubo+13,Umehata+15}. Since an expected average impact parameter between random sight-lines and $z = 3.1$ galaxies in such a high density region should be smaller than that in the normal density region ($\sim 430$\,physical kpc), the $\Delta NB497$ excess at $b \ga 200$\,physical kpc in the SSA22-Sb1 field might be explained by the CGM H~{\sc i} of unobserved galaxies. If the H~{\sc i} absorption enhancement is completely due to the individual galaxies' CGM, the $\Delta NB497$ is likely to correlate with the local overdensity of galaxies, which is, however, inconsistent with our result (Figure~\ref{fig:dM2ovden}). A simple and plausible interpretation of our results is that the extended H~{\sc i} gas, which is different from H~{\sc i} gas clouds associated with each galaxy CGM, distributes only in the SSA22 field. \citet{Yamada+12a} argued that the belt-like LAE overdense structure is an ancestor of local superclusters or ``Great Wall'' rather than a single cluster. In the local superclusters, there is the warm-hot intergalactic medium (WHIM) with $T = 10^{5.5}$ -- $10^7$\,K \citep{Zappacosta+05,Buote+09}. These WHIM is expected to be formed by the shock heating of baryonic matter infalling to the dark matter large-scale structure, where galaxies are also formed. The diffuse H~{\sc i} gas structure revealed by this work may be in the pre-heated phase of the WHIM. 

No correlation between the $\Delta NB497$ and the local LAE overdensity in the SSA22-Sb1 field (Figure~\ref{fig:dM2ovden}) can be explained, at least in the LAE highest density region (LAE overdensity $\ga 3$; see the $4 \times $ the average number density contour in Figure~\ref{fig:dM_skymap_allF}), by a scenario that radiation from galaxies in the overdense region ionizes the gas and reduces the H~{\sc i} amount. Otherwise, hydrogen gas amount is independent of the galaxy local overdensity. \citet{Lee+16} also found a compact (spatial extent $\la 5$\,comoving Mpc) galaxy overdensity structure at $z = 2.3$ with no IGM H~{\sc i} absorption enhancement in their three dimensional tomographic map. They argued that no excess of the H~{\sc i} absorption is due to a lack of the IGM hydrogen gas in/around the galaxy overdensity, while a possibility of suppression of the H~{\sc i} absorption by galaxy feedback or a hot intra-cluster medium (ICM) was also mentioned. They expected that this compact galaxy overdensity evolves to a galaxy group at $z = 0$, not to a cluster. Since the galaxy overdensity found in \citet{Lee+16} and the LAE highest density region in the SSA22-Sb1 field are different in the spatial size and the galaxy population, it is difficult to compare them directly. However, the non-zero H~{\sc i} absorption and large-scale galaxy overdensity in the SSA22-Sb1 field seems to support a picture that the LAE highest density region, which eventually grows into a massive cluster at $z = 0$, contains a significant amount of the hydrogen gas which is partially ionized by the galaxy feedback. 

The size of the H~{\sc i} gas overdensity structure in the SSA22-Sb1 field revealed by this work is $\ga 50$ comoving Mpc because the positive $\Delta NB497$ values are detected over the entire survey area. This size seems to be larger than that of LAE overdensity ($\sim 50$ comoving Mpc from \citealt{Yamada+12a}). The H~{\sc i} structure may generally be extended more than the galaxy structure, because galaxies formed in the middle of the gas structure. The size of the SSA22 H~{\sc i} large-scale structure, $\ga 50$ comoving Mpc, is large, but there are some further larger structures in the lower-$z$ Universe. A traditionally famous large-scale structure is CfA2 Great Wall, which extends at least $\sim 200$\,Mpc \citep{GellerHuchra89}. The largest structure at $z = 0$ is the Sloan Great Wall, which extends over $400$\,Mpc and contains several superclusters consisting of tens of galaxy clusters \citep{Gott+05}. At $z \sim 1$, a larger structure traced by QSOs have been found \citep{Clowes+13}, which reaches $\geq 500$\,Mpc in length. Future applications of our $\Delta NB497$ method to much wider area data are needed to determine the actual extent of the SSA22 super-structure and to reveal the relation between it and these low-$z$ largest structures. If the structure is larger than the upper limit in the scale for cosmological homogeneity (e.g., $370$\,comoving Mpc from \citealt{Yadav+10}), this largest structure ever found in the $z = 3.1$ Universe may challenge the cosmological principle.

\subsection{H~{\sc i} absorption halos around $z = 3.1$ galaxies}

\begin{figure}
\begin{center}
\includegraphics[width=1.0\linewidth, angle=0]{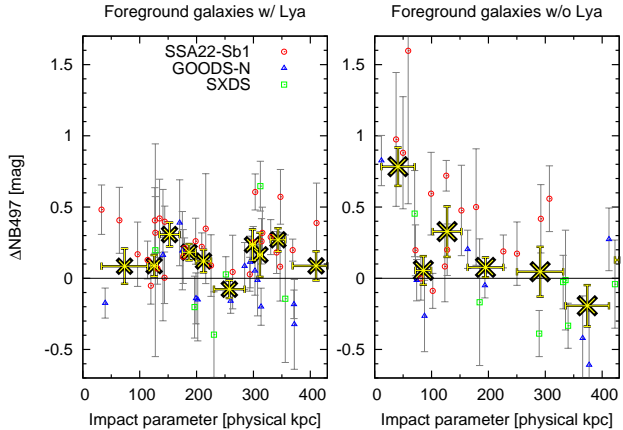}
\caption{Same as Figure~\ref{fig:dM2galdist}, but we divide the sight-line data points by types of the nearest $z = 3.1$ galaxies: left panel for LAEs (${\rm EW^{em}_{rest}} \ga 50$\,\AA) as the foreground and right panel for galaxies without strong Ly$\alpha$ emission (${\rm EW^{em}_{rest}} \la 50$\,\AA) as the foreground. Big yellow crosses correspond to the weighted mean value for every five data points, where we do not divide the sample by the fields. \label{fig:dM2Gdist_Gpop}}
\end{center}
\end{figure}

\begin{figure}
\begin{center}
\includegraphics[width=1.0\linewidth, angle=0]{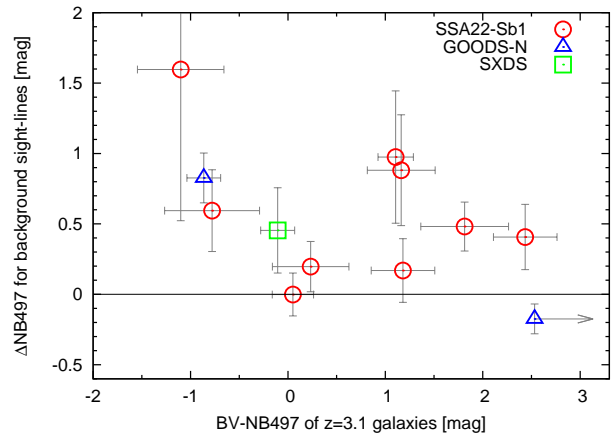}
\caption{$\Delta NB497$ as a function of $BV - NB497$ colour of the nearest $z = 3.1$ object, where only sight-lines with distance from the nearest $z = 3.1$ object less than $100$\,physical kpc are used. \label{fig:BVmNB2Mex}}
\end{center}
\end{figure}

The $\Delta NB497$ estimates are also useful to investigate the radial profile of the H~{\sc i} absorption halo in the galaxy CGM. We found an excess of the $\Delta NB497$ at the impact parameter $< 100$\,physical kpc as shown in Figure~\ref{fig:dM2galdist}, while it is less significant in the control fields. Not only a large scale integration of H~{\sc i} absorption along sight-line (the $NB497$ FWHM corresponds to $\sim 60$\,comoving Mpc at $z = 3.1$) but also different types of galaxies could obscure the correlation between the $\Delta NB497$ and the impact parameter. Especially, LAEs are expected to have different H~{\sc i} gas properties from other populations of galaxies because the hydrogen density, neutrality, and the Ly$\alpha$ photon escape fraction are complicatedly connected each other. We divided the galaxies at $z = 3.1$ into two subsamples: LAEs reported in \citet{Yamada+12a} and other spectroscopically confirmed galaxies. The LAEs of \citet{Yamada+12a} were selected basically by the $BV - NB497$ colour, where the $BV$ is magnitude measured in the $(2 B + V)/3$ composite image. The colour criterion of $BV - NB497 \geq 1$ roughly corresponds to ${\rm EW^{em}_{rest}} \ga 50$\,\AA. Galaxies which are spectroscopically confirmed as $z = 3.1$ but not selected in \citet{Yamada+12a} are expected to have weak or no Ly$\alpha$ emission (${\rm EW^{em}_{rest}} \la 50$\,\AA). In the left panel of Figure~\ref{fig:dM2Gdist_Gpop} we show the $\Delta NB497$ of the background sight-line galaxies for which the nearest $z = 3.1$ galaxies are the LAEs, while those for which the nearest galaxies are the weak Ly$\alpha$ emission galaxies are shown in the right panel. We also show the weighted mean values for every five data points independent of the fields in Figure~\ref{fig:dM2Gdist_Gpop}. In the case that the nearest galaxy have a strong Ly$\alpha$ emission line, we cannot see the increment of the $\Delta NB497$ at $b \la 100$\,physical kpc, while the number of the data points is still limited. A flat $\Delta NB497$ distribution above zero is seen in the SSA22-Sb1 field, which can be due to the large-scale diffuse H~{\sc i} gas unique in the overdensity environment (section~\ref{sec:HI_LSS}). No excess of the $\Delta NB497$ at the impact parameter $< 100$\,physical kpc from the foreground LAEs is in contrast with an excess there for the weak Ly$\alpha$ emission galaxies. This difference suggests that LAEs have thiner H~{\sc i} absorption halos than other star-forming galaxies like LBGs. 

We further investigated the relation between strength of the Ly$\alpha$ emission line from the $z = 3.1$ galaxies and the Ly$\alpha$ absorption by the surrounding H~{\sc i} halos. Figure~\ref{fig:BVmNB2Mex} shows the $\Delta NB497$ as a function of the $BV - NB497$ colour of the nearest $z = 3.1$ galaxy, where we used only the background sight-lines for which the nearest foreground galaxies lies within $b < 100$\,physical kpc. We can see a trend that the H~{\sc i} absorption in the CGM is weaker for foreground galaxies with a larger $BV - NB497$ or stronger Ly$\alpha$ emission. This trend may be due to the small halo mass of the LAEs. Since LAEs have smaller stellar mass ($\sim 10^{6}$ -- $10^9$\,M$_{\odot}$; \citealt{Ono+10}) than continuum-selected galaxies (e.g., LBGs), their halo mass and radius are expected to be smaller in both dark matter and H~{\sc i}. The observed anti-correlation between the $\Delta NB497$ and the $BV - NB497$ of the foreground galaxies may also suggest an anti-correlation between the Ly$\alpha$ escape fraction and the neutrality of the CGM hydrogen gas. Future spectroscopic observations of the CGM metal absorption lines may be complementary because metal absorption have different dependency on the hydrogen neutrality or gas temperature. 

Previous spectroscopic studies for $z \approx 2$ -- $3$ galaxies' CGM showed a clear trend that EW of the H~{\sc i} Ly$\alpha$ absorption decreases with increasing impact parameter, which is well fit by ${\rm EW^{abs}_{rest}} = 0.11 (b/{\rm Mpc})^{-0.76} + 0.25$ \citep{Steidel+10,Rakic+12}. The EWs of the $z = 3.1$ H~{\sc i} absorption can be converted to the $\Delta NB497$ by the analytic form of $\Delta NB497 = -2.5 \log \{1 - {\rm EW^{abs}_{rest}}\times(1+3.1)/78$\AA$\}$. This conversion formula should be correct if the wavelength range used for calculation of the absorption in the EW and $\Delta NB497$ estimations is same, as shown in Figure~\ref{fig:dM2specEW}. Even if the absorption width assumed in the spectroscopic EW estimation is narrower than that in the photometric $\Delta NB497$ estimation, which is a typical case, the conversion formula can be adopted in general fields where galaxies and H~{\sc i} gas distribute randomly along a line-of-sight. We superpose the $\Delta NB497$ estimated from the $b$ -- EW relation of \citet{Rakic+12} on our own measurements in Figure~\ref{fig:dM2galdist}. We have assumed no evolution of the CGM properties from $z \sim 2.4$ where \citet{Rakic+12} investigated to $z = 3.1$. The $\Delta NB497$ expected from \citet{Rakic+12} is less than $0.1$\,mag at $b \ga 50$\,physical kpc, which may be consistent with our $\Delta NB497$ distributions around zero with large uncertainties in the two control fields. On the other hand, the $\Delta NB497$ values in the SSA22-Sb1 field seem to be larger than those expected from \citet{Rakic+12}, at least by a factor of $5$ at almost all impact parameters. Even if we add $0.2$\,mag to the $\Delta NB497$ of \citet{Rakic+12} taking account of the diffuse IGM H~{\sc i} component unique in the SSA22 field, there still seems to be a $\Delta NB497$ excess at $b \la 100$\,physical kpc relative to the previous studies. This suggests that larger amount of H~{\sc i} gas lies not only in the IGM but also in the CGM in the SSA22 proto-cluster than in the general region. A significant environmental dependency of the scale length of the Ly$\alpha$ emitting halos associated with the LBGs and LAEs at $2 \la z \la 3$ has been reported \citep{Matsuda+12, Momose+16}, where the scale length in the proto-cluster environment is larger than that in the general fields by a factor of $\sim 3$. If the CGM H~{\sc i} absorption halos have the similar trend, it may explain the $\Delta NB497$ enhancement in the SSA22 field relative to the previous studies, which should be confirmed by future spectroscopic follow-up observations.

\section{Summary}

In this work, we developed a new scheme, the ``$\Delta NB497$'' method, to characterize strength of Ly$\alpha$ absorption by the $z = 3.1$ H~{\sc i} gas photometrically. We applied the scheme to the multi-band photometry in the SSA22-Sb1 field and the two control fields after a careful calibration of photometric colours and a careful selection of the background sight-line galaxies. Our main results are as follows. 
\begin{itemize}
\item The $\Delta NB497$ estimates are reasonably correlated with the absorption EWs measured using high S/N spectra, which supports the reliability of our scheme.
\item The $\Delta NB497$ values in the SSA22-Sb1 field are significantly larger than those measured in the two independent control fields.
\item The $\Delta NB497$ enhancement is detected over the entire SSA22-Sb1 field, suggesting that the H~{\sc i} overdensity structure spreads out $\ga 50$\,comoving Mpc. The size is larger than the LAE overdensity structure ($\sim 50$\,comoving Mpc from \citealt{Yamada+12a}). 
\item No clear dependency of the $\Delta NB497$ on the local LAE overdensity in a few comoving Mpc scale was found, while the field-averaged ($\sim 50$\,comoving Mpc scale) quantities are correlated. 
\item In the SSA22-Sb1 field, a significant $\Delta NB497$ excess was detected even at a large distance from the nearest $z = 3.1$ galaxies (impact parameter $b \ga 200$\,physical kpc). 
\item There seems to be the diffuse and large-scale H~{\sc i} gas component lying in/around the SSA22 proto-cluster, which is independent of the individual galaxy CGM. 
\item An excess of the $\Delta NB497$ at the impact parameter $b \la 100$\,physical kpc was found at least in the SSA22-Sb1 field, which is considered to be due to the individual CGM H~{\sc i} gas. 
\item The H~{\sc i} absorption is weaker in the CGM of strong LAEs, suggesting that the Ly$\alpha$ escape fraction from galaxies depends on hydrogen neutrality in the CGM. 
\item The $\Delta NB497$ values at $b < 100$\,physical kpc in the SSA22-Sb1 field are larger than those expected from the previous studies even after taking account of the diffuse H~{\sc i} component in the IGM, which implies that the CGM absorption halos have the larger scale length in the porto-cluster environment.   
\end{itemize}

Wider-field multi-band photometry is needed to determine the actual extent of the H~{\sc i} overdensity structure and to investigate relationship between the SSA22 super-structure and the local largest structure (e.g., Great Wall). Spectroscopic observations for the IGM metal absorptions using next-generation instruments (Subaru/PFS or TMT) will help us to understand physical mechanism responsible for the difference of the H~{\sc i} absorption strength depending on the foreground galaxy types.

\section*{Acknowledgements}
K. M. and A. K. I. are financially supported by JSPS KAKENHI Grant Number 26287034. H. U. is supported by JSPS Grant-in-Aid for Research Activity Start-up (16H06713). 












\bsp	
\label{lastpage}
\end{document}